\documentclass[a4paper,11pt]{article}
\usepackage{jinstpub} 
\usepackage{lineno}
\usepackage{makecell}
\usepackage{siunitx}   
\usepackage{subcaption}   

\newcommand{\um}{\si{\micro\metre}}
\newcommand{\urad}{\si{\micro\radian}}

\title{Test beam measurements and computer simulations of the ATLAS ITk R2 silicon strip detector}

\author[a]{J.-H.~Arling,}
\author[b]{P.~Federic,}
\author[a]{Y.~He,}
\author[c]{C.~M.~Helling,}
\author[d]{N.~Hessey,}
\author[a]{L.~Huth,}
\author[d]{G.~Jain,}
\author[e]{C.~Jessiman,}
\author[d,f]{S.~Manson,}
\author[d,f]{L.~Poley,}
\author[e]{J.~S.~Keller,}
\author[g]{J.~Kroll,}
\author[g]{J.~Kvasnicka,}
\author[g,h,1]{R.~Privara,\note{Corresponding author.}}
\author[d,f]{P.~Speers,}
\author[g]{P.~Tuma.}

\affiliation[a]{Deutsches Elektronen-Synchrotron DESY,\\
Notkestraße 85, Hamburg, 22607, Germany}
\affiliation[b]{Faculty of Mathematics and Physics, Charles University,\\
Ke Karlovu 2027/3, Prague 2, 121 16, Czech Republic}
\affiliation[c]{Department of Physics \& Astronomy, University of British Columbia,\\
325 - 6224 Agricultural Road, Vancouver, BC V6T 1Z1, Canada}
\affiliation[d]{TRIUMF,\\
4004 Wesbrook Mall, Vancouver, BC V6T 2A3, Canada}
\affiliation[e]{Department of Physics, Carleton University,\\
1125 Colonel By Drive, Ottawa, ON K1S 5B6, Canada}
\affiliation[f]{Department of Physics, Simon Fraser University,\\
University Dr W, Burnaby, BC V5A 1S6, Canada}
\affiliation[g]{Institute of Physics, Czech Academy of Sciences,\\
Na Slovance 1999/2, Prague 8, 182 00, Czech Republic}
\affiliation[h]{Faculty of Science, Palacky University in Olomouc,\\
17. listopadu 1192/12, Olomouc, 779 00, Czech Republic}

\emailAdd{radek.privara@cern.ch}

\abstract{
The ATLAS Inner Tracker, the future innermost part of the ATLAS detector, is an all-silicon tracker composed of pixel and strip modules, designed to cope with the extreme conditions expected during High-Luminosity LHC runs. 
Thorough testing of modules during the individual phases of their development is critical to ensure the required performance level of the whole tracker. 
This document presents results obtained from electron beam measurements of the ATLAS ITk R2 end-cap strip module. 
Key performance metrics are presented and discussed for both perpendicular and angled beam incidence.
Computer simulations of the module were performed in the \texttt{Allpix-Squared} framework and the results were compared to experimental data.}

\keywords{Performance of High Energy Physics Detectors, Particle tracking detectors, Si microstrip and pad detectors, Detector modelling and simulations II.}

\begin{document}
\maketitle
\flushbottom

\section{Introduction}
The upcoming High-Luminosity Large Hadron Collider (HL-LHC) will deliver 5--7.5 times higher instantaneous luminosity when compared to the LHC nominal value, with up to 200 inelastic proton-proton collisions per bunch crossing \cite{ITk-TDR}. 
To cope with these conditions while ensuring a sufficient tracking performance level of the ATLAS experiment, a new all-silicon tracker with increased accuracy, granularity and radiation tolerance is being built.
The ATLAS Inner Tracker (ITk) will consist of pixel modules positioned close to the beam line, totaling 13 m$^2$ of surface area and 5 billion readout channels, and of strip modules further away, covering 165~m$^2$ with a total of 60 million channels \cite{ITk-TDR}.

Pre-production ITk strip modules are tested using charged particle beams to ensure their full functionality and sufficient performance level. 
Additionally, testing of modules irradiated by protons, neutrons and gamma particles is done to evaluate their performance at or beyond their expected end-of-life state.
Two performance criteria must be simultaneously satisfied for a range of charge thresholds (also called an operating window) on any given strip module and at any point of its lifetime: detection efficiency above 99 \% and noise occupancy below 0.1 \%. 
These requirements are equivalent to establishing a signal-to-noise ratio of at least 10~\cite{ITk-TDR}.

Several early prototypes of ITk strip modules have been studied in the past and have shown satisfactory performance after irradiation \cite{TB_2020_NIMA}.
Analysis of more mature, pre-production stage strip modules has been ongoing \cite{JINST_SS_temp-ref}, likewise with positive results.

This document presents analysis of charged particle measurements of a pre-production ATLAS ITk R2 strip module.
Performance of the module has been studied for both perpendicular and non-perpendicular incidence of the particle beam.
This study is motivated by the ATLAS detector magnetic field which will cause particles to pass through the ITk modules at various incidence angles based on their transverse momentum and the position of a given module. 

Computer simulations are used to complement the measurements, recreating the experimental setup with the particle beam in the \texttt{Allpix-Squared} framework~\cite{APSQ-ref}, recently extended by the ITk collaboration to enable simulation of all types of ITk strip modules.
Simulations can be used to analyze the proposed experimental setup or predict module performance, provided they have been validated and shown to be matching the experimental data well.
Comparison of simulation and experimental results is presented in this document with the goal of establishing the level of agreement and identifying potential shortcomings of the current implementation of strip modules in the simulation framework.

\section{ATLAS ITk Strip}
\begin{figure}[hbtp]
    \centering
    \includegraphics[width=0.98\textwidth]{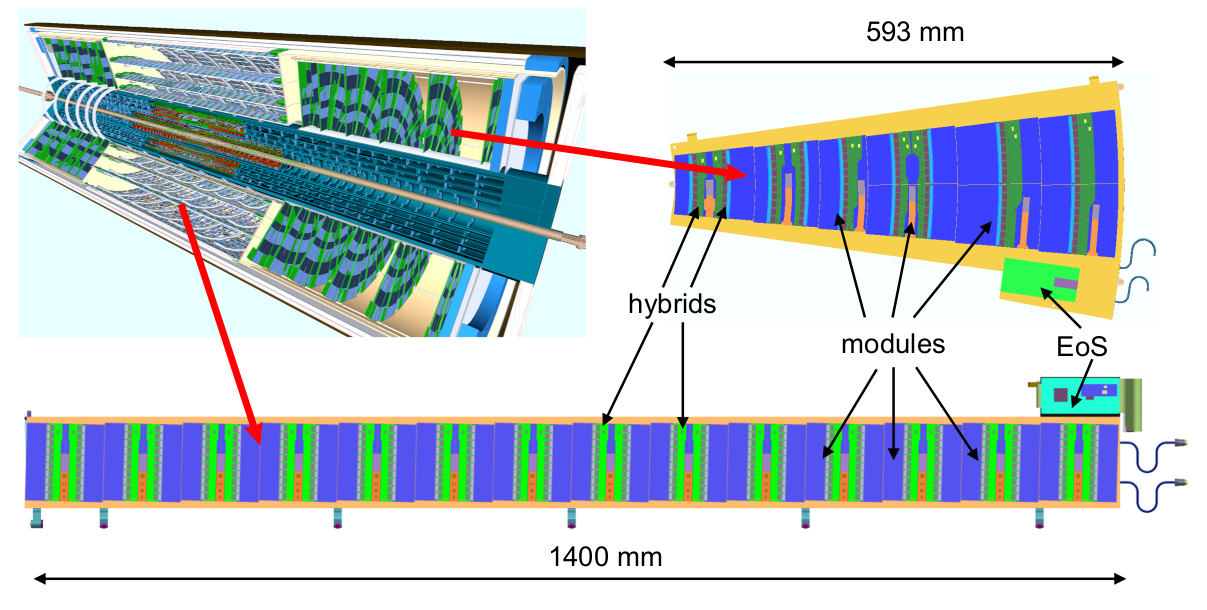}
    \caption{The layout of the ATLAS ITk (top left). In the forward regions, the trapezoidal end-cap strip modules are laid out on "petals" (top right) and in the barrel section the rectangular modules are placed on "staves" (bottom). From \cite{ITk-TDR}.}
    \label{fig:itk_layout}
\end{figure}

The layout of the ITk strip detector, shown in Figure~\ref{fig:itk_layout}, is separated into a central barrel section, consisting of four concentric cylindrical layers of barrel strip modules assembled on so-called stave structures, and two end-cap sections in the forward regions, each formed of six disks of end-cap modules laid out on structures called petals.
A total of eight types of strip modules will be installed in the ITk, two types of rectangular barrel modules (Short-Strip and Long-Strip) with parallel strips and six types of end-cap modules (designated as R0--R5) featuring a trapezoidal sensor with radial strip geometry.

\begin{figure}[hbtp]
    \centering
    \includegraphics[width=0.65\textwidth]{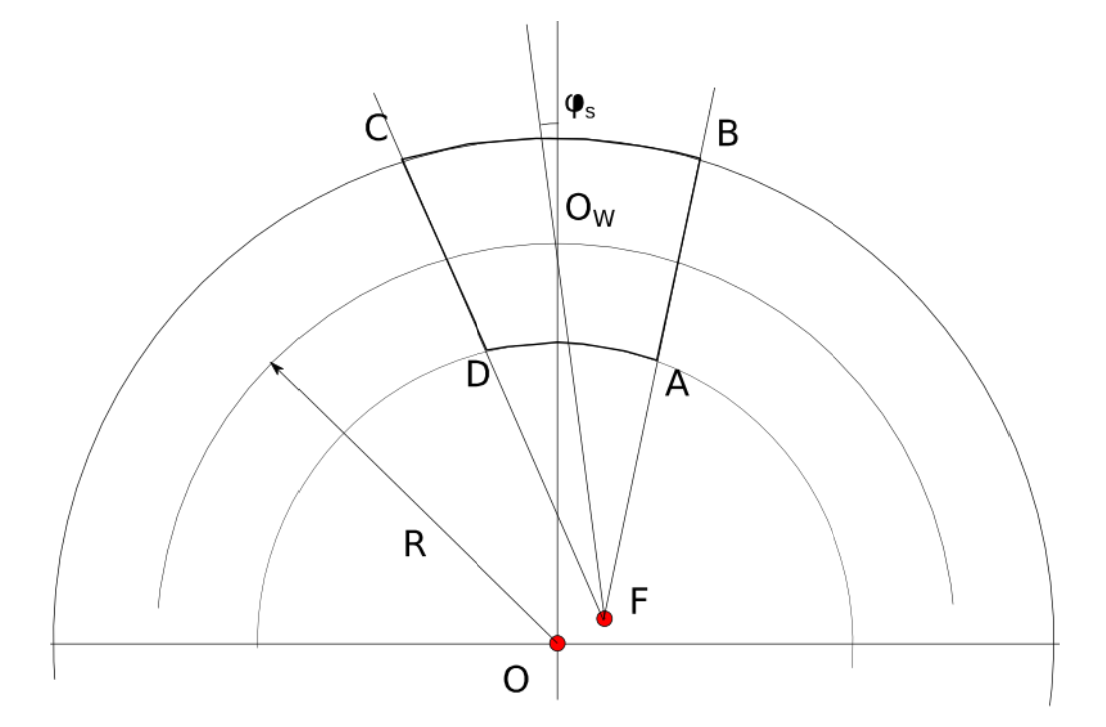}
    \caption{The illustration of the stereo annulus strip sensor geometry featuring the stereo angle $\varphi_s$. The sensor shape is defined by the A, B, C, D points. The end-cap disk centre is marked by O and the focal point for sensor sides and strips by F. From \cite{ITk-TDR}.}
    \label{fig:stereo_angle}
\end{figure}

The specific sensor design (and the associated sensor shape) of end-cap strip sensors is referred to as stereo annulus \cite{StereoAnnulus-ref}.
The inner and outer edges of a stereo annulus strip sensor are concentric arcs centered at the centre of the end-cap disk, while the sensor sides (and the strips) converge at a focal point displaced from the disk centre by the stereo angle, as is shown in Figure~\ref{fig:stereo_angle}.
This feature enables 3D space point reconstruction of the particle hit when two end-cap modules are placed back-to-back, as will be the case for all strip modules in the ITk.
The stereo angle is 20 mrad for all ITk end-cap strip modules~\cite{ITk-TDR}.

\begin{figure}[hbtp]
    \centering
    \includegraphics[width=0.65\textwidth]{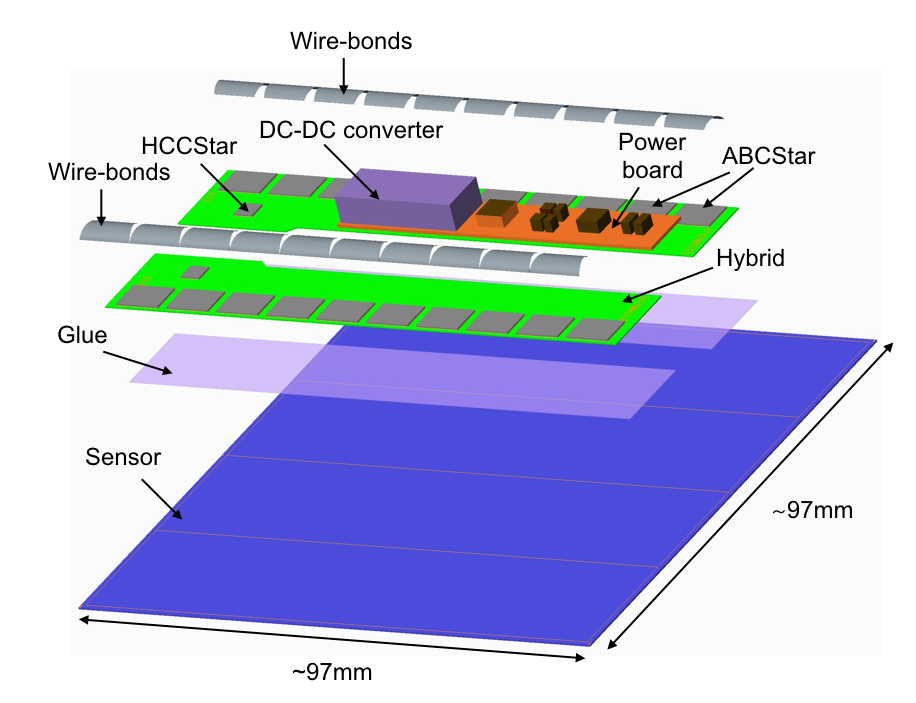}
    \caption{Components of an ITk Short-Strip barrel module featuring a rectangular sensor with four rows of strips, two hybrids and a single power board. ITk end-cap strip modules feature the same component groups. From \cite{ITk-TDR}.}
    \label{fig:itk_components}
\end{figure}

The sensitive part of all ITk strip module types is an n$^+$-in-p float-zone silicon sensor with the physical thickness of 320 \um\ and the strip pitch of 75.5 \um\ for barrel modules and between 69 and 84~\um\ for end-cap modules \cite{ITk-TDR}.
The sensors are divided into several segments which are read out as separate data streams.
Printed-circuit boards called hybrids are glued directly on the sensor, providing mounting surface for other components such as front-end readout ASICs (ABCStar~\cite{ABCStar-ref}) and data aggregator chips (HCCStar~\cite{HCCStar-ref}). 
The front-end channels are AC-coupled to the n-type strips. Additionally, a power board is glued on the sensor, housing a DC-DC power converter \cite{Faccio:2020DV}, an Autonomous Monitor and Control Chip (AMAC) \cite{Gosart_2023} and a high-voltage GaN FET switch. 
The components of a Short-Strip barrel module are shown in Figure~\ref{fig:itk_components}. 
While barrel and end-cap module types differ in size and shape, they feature equivalent design and utilize identical component groups.
The modules are operated in a binary readout mode where a hit on a given strip is registered if the charge collected on that strip surpasses a set charge threshold.

\section{Device-under-test}
The measurements presented herein were performed at the DESY II test beam facility~\cite{DESY-ref} in June~2022 with an unirradiated, pre-production ITk R2 module, operated at 500~V of reverse-bias voltage and therefore in full depletion.

\begin{figure}[htbp]		
	\centering
	\includegraphics[width=0.75\textwidth]{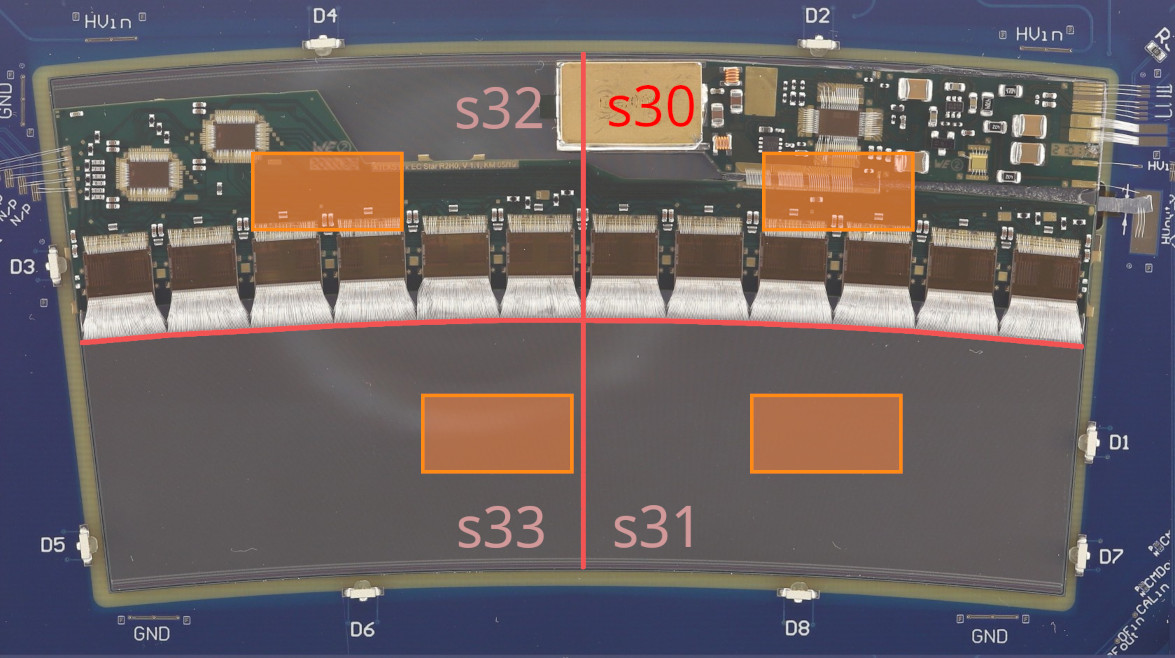}
	\caption{The ITk R2 end-cap module featuring two strip rows, a hybrid and a power board. The module is mounted on a test circuit board. Red lines delineate the four individual module segments s30 to s33, which are denoted in the photo. Orange rectangles approximately mark out the beam spot location during the measurements of each segments.}
	\label{fig:dut}
\end{figure}

The ITk R2 module is shown in Figure~\ref{fig:dut}.
It features two rows of 1538 strips 30.8~mm long, with the strip pitch of 74.5 \um\ in the lower strip row and 78.4 \um\ in the upper strip row.
The angular strip pitch is 129.08 \urad\ in both rows \cite{ITk-TDR}.
The module features a single hybrid and a power board, both glued onto the upper strip row.
Data is read out in four data streams corresponding to the four module segments -- named s30 and s32 in the upper row, s31 and s33 in the lower row.

The measured module, also referred to as the device-under-test (DUT), was tested by performing threshold scans which consist of several tens of repeated measurements (runs) with a progressively increasing charge threshold above which a hit is registered in the DUT. 
For this reason, performance metrics are generally presented as a function of the charge threshold.
Approximately 250 thousand events were taken in every run. 

One threshold scan was dedicated to each of the four segments of the DUT to establish its overall performance level. 
Additional scans were taken with the beam impacting the same general area of the segment s33 at various angles of incidence.
This was achieved by rotating the module by 5$^\circ$ to $25^\circ$ around the vertical axis, which lies in the sensor plane and coincides with the central strip direction.
This rotation causes the particles to pass through the sensor at an angle, prolonging the particle's path in the sensor material and increasing the amount of deposited charge, but also leading to the charge being split among multiple strips.
The amount of signal on the leading strip (the strip with the most charge) is therefore reduced.

Rotations around the horizontal axis, which lies in the sensor plane and is perpendicular to the central strip direction, were not performed.
Such a rotation also increases the amount of deposited charge, but does not cause the charge to be split among more strips.
Zero rotation around the horizontal axis thus represents the worst case scenario and measuring under such conditions is desirable when determining the module performance.

\section{Experimental setup}
The DESY II test beam facility provides a beam of electrons with user-configurable momenta of 1--6~GeV/c \cite{DESY-ref}.
For this analysis, the beam was tuned for the momentum of 5 GeV/c.
Tracking of beam particles was ensured by the DURANTA EUDET-type telescope \cite{EUDET-ref} equipped with six Mimosa26 pixel detectors \cite{Mimosa26-ref}, each totaling $1152\times 576$ pixels with the pitch of $18.4 \times 18.4$~\um$^2$. 
The Mimosa26 detectors use a rolling-shutter read-out with an integration time of 115.2 $\si{\micro\second}$, which allows multiple electrons to arrive within the same read-out window.
To resolve this potential ambiguity in the arrival times of individual particles, an FE-I4 pixel detector~\cite{FEI4-ref} composed of $80 \times336$~pixels with the pitch of $250\times 50$ \um$^2$ and a read-out window of 25 ns was used for the efficiency evaluation of the DUT.

The primary trigger signal was generated by two pairs of scintillators positioned in the beam, one in front of and one behind the telescope, requiring the activation of all four scintillators to issue the trigger signal.
An AIDA Trigger Logic Unit (TLU) \cite{TLU-ref} was then used for distributing the trigger signal to all detectors in the setup.
Communication with the individual detectors connected into the experimental setup, as well as collection and storage of data measured by the detectors, were handled by the \texttt{EUDAQ2} framework~\cite{EUDAQ2-ref}.

During the measurement, the DUT was placed in a styrofoam cooling box positioned in the centre of the telescope, between the third and fourth telescope planes.
The box features metal cylinder inserts for dry ice cooling, making it possible to reach air temperatures below -40~$^\circ$C in the box.
The box was additionally mounted on a moving stage allowing for precise positioning and rotation of the box with respect to the beam.

\section{Simulation}
Test beam measurements were complemented with simulations performed in the \texttt{Allpix-Squared} framework. 
Rectangular strip detectors can be simulated with ease in the framework and since the release version 3.0.0, the framework has been extended by the ITk strip collaboration to allow simulation of end-cap strip detectors such as the ITk R2, faithfully reproducing the complex radial strip geometry and properly implementing the stereo angle feature into the detector model.

The framework utilizes the Geant4 toolkit~\cite{G4-ref} to simulate the passage of particles through physical volumes and the deposition of energy therein.
The deposited energy is converted to charge carriers which are propagated in the sensor volume based on the provided electric field.
The propagated charge is then collected on the readout side of the detector and assigned to individual pixels or strips, depending on the detector type.
Gaussian noise is applied to these collected charges.
Noise is however not applied to pixels or strips where no charge has been collected, meaning hits caused purely by noise cannot be generated in this approach.

The arrangement of detectors in the simulations was identical to the test beam measurements and geometrically accurate models of the individual detectors were used.
A simple model of the DUT cooling box was created, consisting only of the styrofoam layers of the box.
A linear electric field was simulated inside the DUT sensor.
The DUT electronics noise was set to 0.106 fC, or 662~$e^-$~ENC (Equivalent Noise Charge), the average input noise value across the DUT channels.
The input noise values for individual channels were obtained by scanning the charge threshold parameter of the DUT without the electron beam, using the internal calibration circuit on the module to inject a fixed charge into the front-end amplifier channels \cite{ITk_FE-ref}.

Cross talk was estimated using a capacitance-based model \cite{strip-crosstalk}.
It was simulated by removing 1~\% of the total collected charge as a result of the capacitive coupling between the strips and the sensor backplane. 
Additionally, 2.8 \% of the total charge collected on a given strip was transferred to the adjacent strips (1.4 \% to each) as a result of the inter-strip capacitive couplings. 
These fractions are based on coupling, bulk and inter-strip capacitance values measured on an ITk R2 sensor.

\section{Data reconstruction and analysis}
Reconstruction and analysis of the collected data was performed in the \texttt{Corryvreckan} framework~\cite{Corry-ref} which can properly account for the radial geometry of the ITk strip end-cap modules. 
Simulation data obtained from the \texttt{Allpix-Squared} framework were likewise imported into \texttt{Corryvreckan} and analyzed identically to test beam data.

During the test beam measurements, the charge threshold of the DUT is set in internal units of the front-end electronics called DAC.
The conversion from DAC to a common unit of fC is done in two steps.
First, the conversion of DAC to voltage is done via a lookup table shared by all ITk strip modules.
The subsequent conversion from V to fC is then performed for each individual read-out ASIC using conversion coefficients obtained from built-in module tests, where a fixed charge is injected into the read-out channels.

Particle tracks were reconstructed based on hit clusters (collections of adjacent pixels or strips with a hit) in the telescope detectors and the FE-I4 only.
The DUT was excluded from the tracking process.
Each detector was required to have a hit cluster present within a certain distance, called a spatial cut, from the intersection with the track candidate. 
The tracks were then used to align the telescope detectors and the FE-I4 by minimizing the track-fit $\chi^2$.
The spatial cuts were progressively tightened in several iterations of the alignment process, with the final iteration requiring hits found within 100~\um\ from the track intersection with a given detector.

The subsequent DUT alignment was performed by moving the DUT in the plane perpendicular to the beam and thus minimizing the unbiased residuals, the distance between the DUT hit position and the intersection of the reconstructed track with the DUT.
The DUT alignment was likewise done iteratively, with progressively tighter spatial cuts defined in polar coordinates due to the end-cap module geometry. 
The final DUT alignment iteration used 320~\urad\ as the angular spatial cut value, equivalent to 2.5 times the ITk R2 angular strip pitch.

The detection efficiency for a given charge threshold was defined as a fraction whose numerator is the number of tracks which exhibit a hit within the spatial cuts in the DUT and the denominator is the number of all tracks. 
The efficiency evaluation was also done using the 320~\urad\ angular spatial cut while additionally selecting only high-quality tracks with the value of the track-fit $\chi^2$/ndf parameter lower than 2.

Efficiency points were fitted with an empirical skewed complementary error function
\begin{equation}
    f(x)=\frac{1}{2}\cdot\epsilon_\text{max}\cdot\text{erfc}\left[\frac{x-q_{50}}{\sqrt{2}\sigma}\cdot\left(1-0.6\tanh\left(\frac{A(x-q_{50})}{\sqrt{2}\sigma}\right)\right)\right]\, ,
    \label{eq:eff_fit}
\end{equation}
where $\text{erfc}$ denotes the complementary error function, $\tanh$ the hyperbolic tangent function, $\epsilon_\text{max}$ is the maximum efficiency, $\sigma$ is the width of the error function distribution and $A$ is a skew parameter~\cite{eff_fit-ref}. 
The parameter $q_{50}$ is the charge threshold where exactly 50 \% of the tracks exhibit a hit in the DUT and it corresponds to the median of the collected charge distribution on the leading strip~\cite{TB_2020_NIMA}.
For signal-to-noise evaluation, the $q_{50}$ represents the amount of signal.

The input noise values used for signal-to-noise evaluation were obtained by averaging input noise over channels (strips) in the beam impact area.
Noise occupancy, a related performance metric describing the rate of observed hits caused by noise only, was obtained by scanning the charge threshold without the beam present and without injecting any charge into the read-out channels.

\section{Results}
\subsection{Segment scans}
The median charge on the leading strip, obtained for the individual DUT segments by fitting the efficiency data points, ranges from $(3.65\pm0.16)$ fC to $(3.81\pm0.34)$ fC.
Sources of the uncertainty include the uncertainty introduced by the fit function, the systematic variance of individual strips and the statistical uncertainty of the median charge value~\cite{JINST_SS_temp-ref}.
The strip variance is evaluated by analyzing the efficiency of individual strips, thus obtaining the median charge for each strip from a fit.
The standard deviation of this median charge distribution is then interpreted as the systematic uncertainty which accounts for a number of effects: read-out ASIC manufacturing variations, differences in the calibration of individual strips and capacitive effects from components placed on the sensor.
The statistical uncertainty of the median charge is obtained as the standard error of the mean of the median charge distribution.
The uncertainty components are listed for each DUT segment in Table~\ref{tab:seg_syst}, showing that the systematic strip variance is the dominant contribution.

\begin{table}[hbtp]
    \centering
    \caption{Median charge uncertainty components and the total uncertainty for each DUT segment.}
    \begin{tabular}{c|c|c|c||c} 
        \hline
        Segment & Fit [fC] & Strip systematics [fC] & Statistical [fC] & Total [fC] \\ \hline
        s30 & 0.007 & 0.340 & 0.024 & 0.341 \\
        s31 & 0.005 & 0.343 & 0.023 & 0.344 \\
        s32 & 0.007 & 0.160 & 0.011 & 0.161 \\
        s33 & 0.006 & 0.242 & 0.016 & 0.243 \\ \hline
    \end{tabular}
    \label{tab:seg_syst}
\end{table}

Mean input noise values for the DUT segments cover the range of $(0.1011\pm0.0007)$~fC to $(0.1104\pm0.0008)$~fC.
The uncertainty of the presented noise values is statistical and equal to the standard deviation of values from which the mean is calculated.
The resulting signal-to-noise ratios range from $33.0\pm1.5$ on the s32 segment to $37.7\pm3.4$ on the s31 segment.
The values for each segment are mutually consistent within their respective uncertainties.

As expected for an unirradiated strip module, the obtained signal-to-noise values are several times higher than the collaboration requirement for modules at their end-of-life state.
For all DUT segments, a wide window of charge thresholds satisfying the requirements on detection efficiency and noise occupancy is found, with widths ranging from 1.28 to 1.48~fC. 

\begin{figure}[htbp]
    \begin{subfigure}{0.49\textwidth}
        \centering
        \includegraphics[width=1\textwidth]{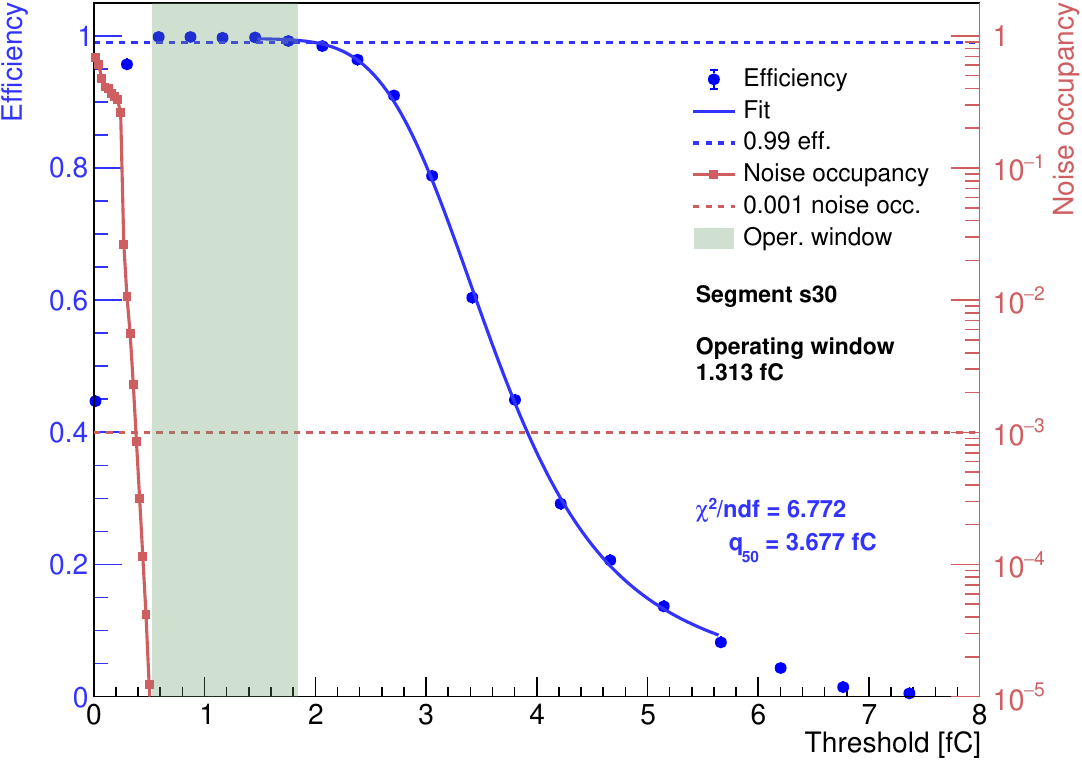}
        \caption{DUT segment s30.}
    \end{subfigure}
    \begin{subfigure}{0.49\textwidth}
        \centering
        \includegraphics[width=1\textwidth]{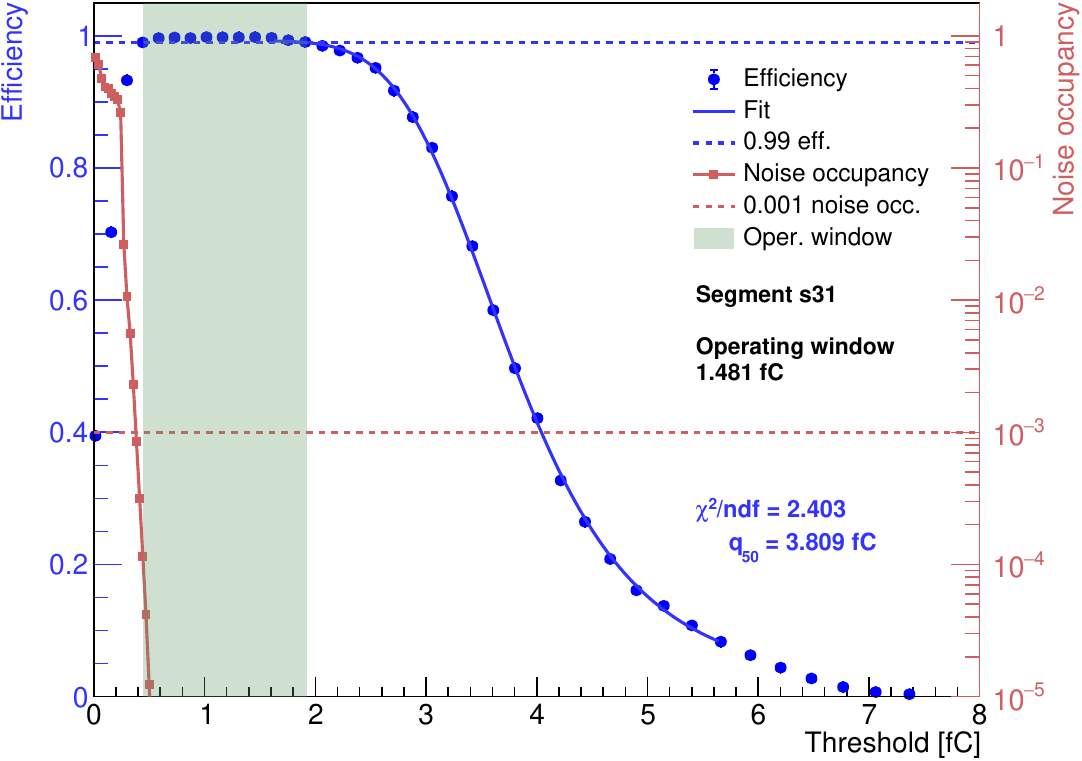}
        \caption{DUT segment s31.}
    \end{subfigure}\vskip0.6cm
    \begin{subfigure}{0.49\textwidth}
        \centering
        \includegraphics[width=1\textwidth]{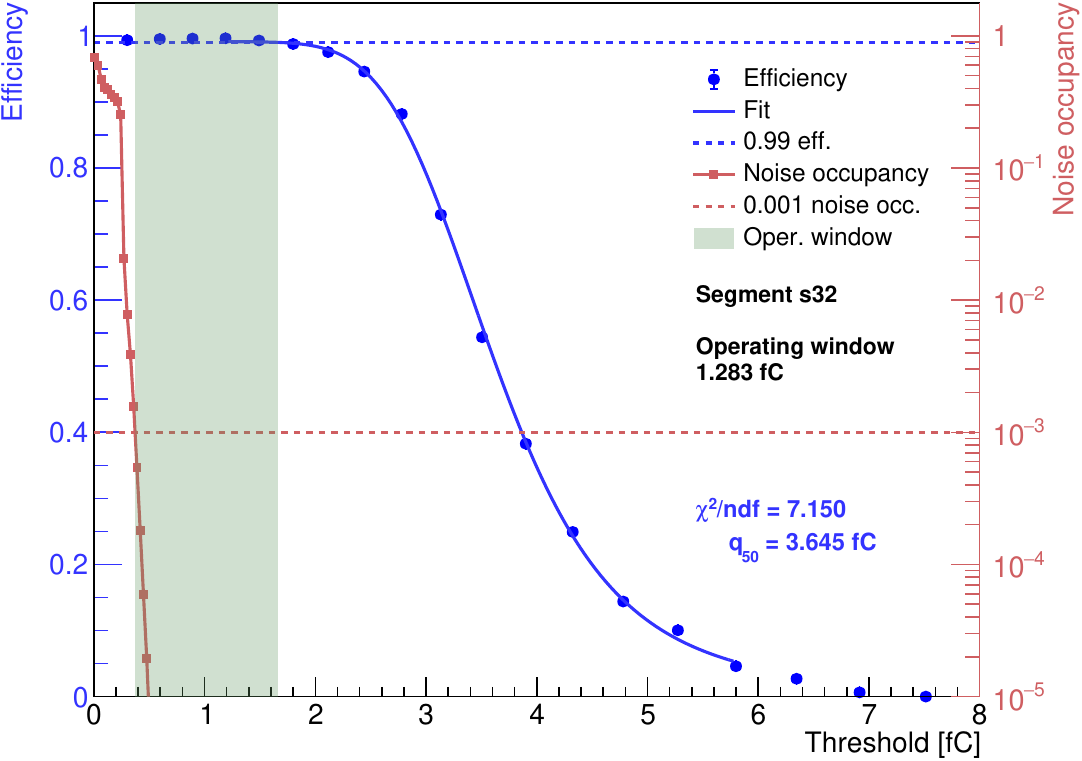}
        \caption{DUT segment s32.}
    \end{subfigure}
    \begin{subfigure}{0.49\textwidth}
        \centering
        \includegraphics[width=1\textwidth]{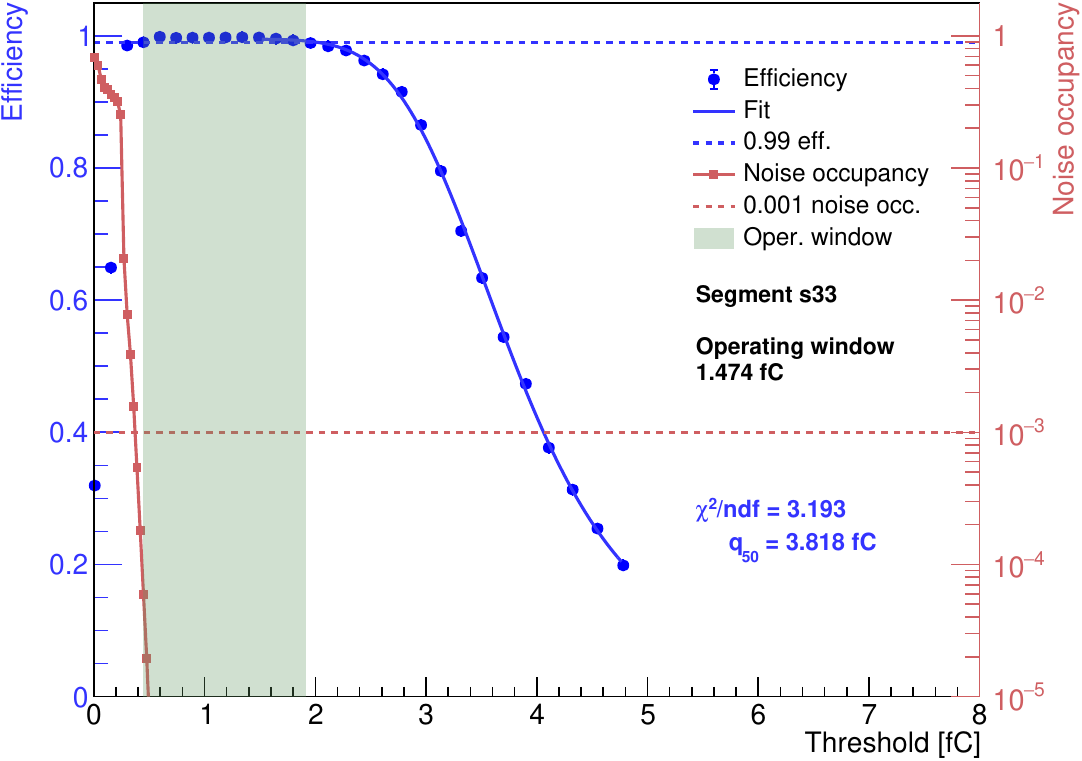}
        \caption{DUT segment s33.}
    \end{subfigure}
	\caption{Detection efficiency (blue) and noise occupancy (red) as functions of the charge threshold for all DUT segments. Efficiency points are fitted using Function~\ref{eq:eff_fit} up to approximately 5.5 fC as including the lower efficiency points worsens the fit quality in the critical 50 \% efficiency region. The dashed blue line denotes the required 99\% efficiency level and the dashed red line the required 0.1\% noise occupancy level. The pale green region denotes the operating window. The DUT is inefficient at the lowest charge thresholds due to noise hits pulling the cluster centre away from the particle hit and therefore failing the spatial cut requirement.}
	\label{fig:segment_eff_individual}
\end{figure}

Segments s30 and s32, where the hybrid and the power board are positioned, exhibit similar performance in terms of the leading strip median charge and the operating window width. 
Likewise, segments s31 and s33 behave similarly to one another, showing higher leading strip median charge and wider operating windows than the segments beneath the hybrid and power board.
The efficiency and noise occupancy curves for all DUT segments, as well as the operating windows, are shown in Figure~\ref{fig:segment_eff_individual}.
The performance metrics are listed in Table~\ref{tab:segments}. 

\begin{table}[hbtp]
    \centering
    \caption{Performance parameters of all four DUT segments for the perpendicular incidence of the electron beam.}
    \begin{tabular}{c|c|c|c|c} 
        \hline
        Segment & \makecell{Median \\ charge [fC]} & Input noise [fC] & S/N & \makecell{Operating window\\ width [fC]} \\ \hline
        s30 & $3.68 \pm 0.34$ & $0.1042 \pm 0.0009$ & $35.3 \pm 3.3$ & 1.31 \\
        s31 & $3.81 \pm 0.34$ & $0.1011 \pm 0.0007$ & $37.7 \pm 3.4$ & 1.48 \\
        s32 & $3.65 \pm 0.16$ & $0.1104 \pm 0.0008$ & $33.0 \pm 1.5$ & 1.28 \\
        s33 & $3.82 \pm 0.24$ & $0.1064 \pm 0.0007$ & $35.9 \pm 2.3$ & 1.47 \\ \hline
    \end{tabular}
    \label{tab:segments}
\end{table}

\begin{figure}[htbp]
	\centering
	\includegraphics[width=0.48\textwidth]{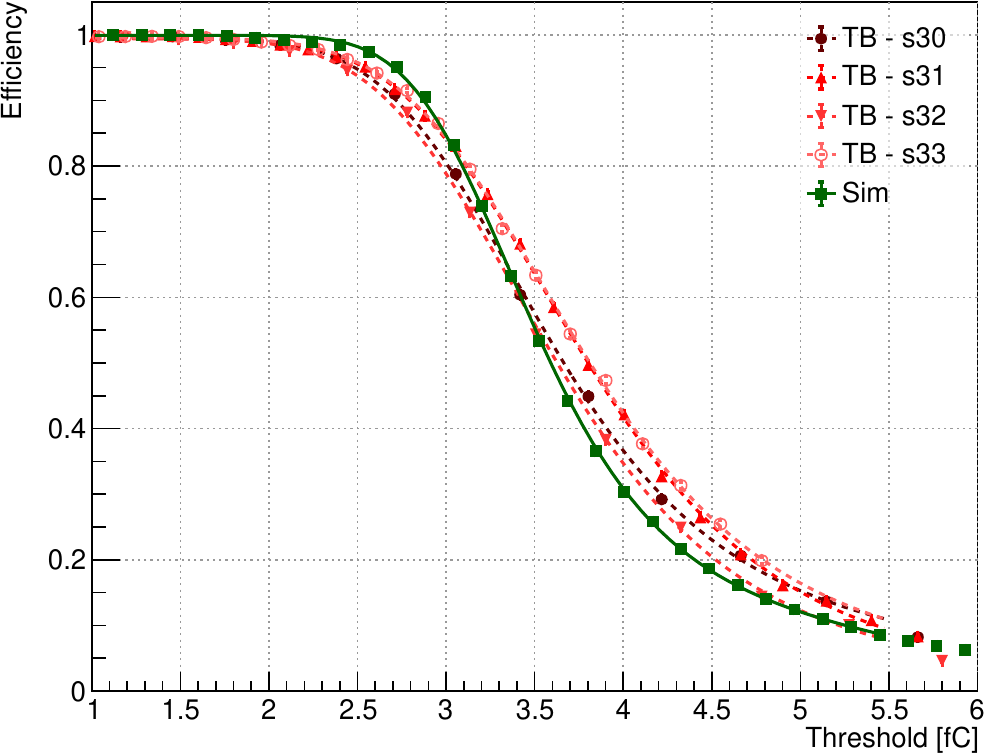}\hskip0.5cm
 	\includegraphics[width=0.48\textwidth]{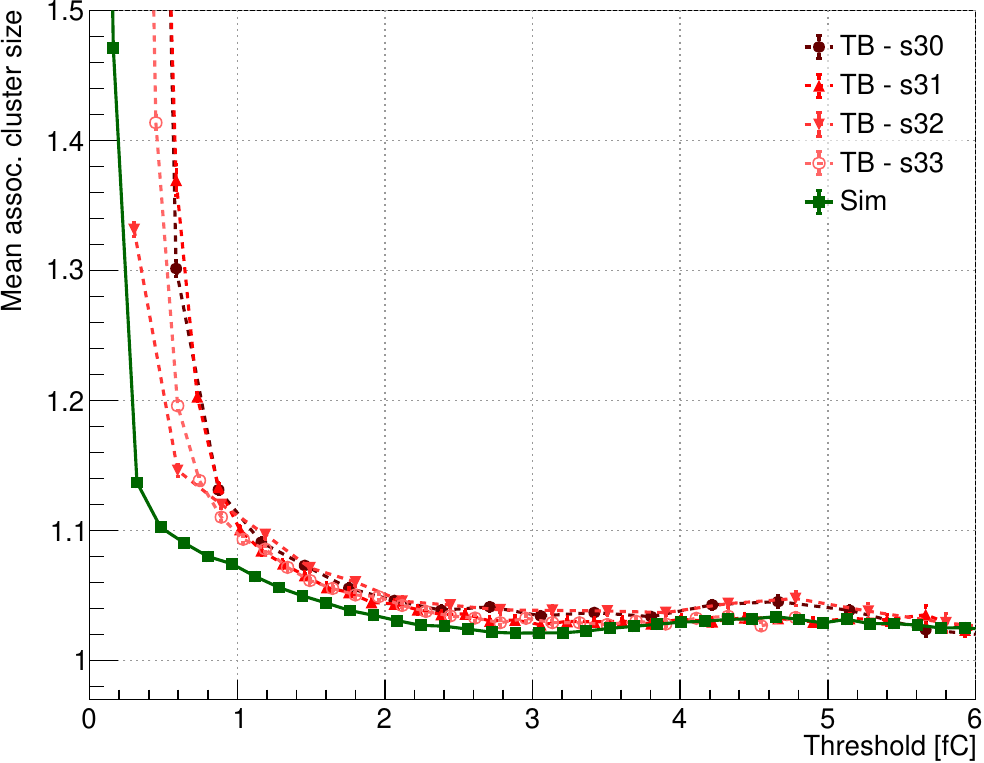}
	\caption{Detection efficiency (left) and mean associated cluster size (right) as a function of the charge threshold obtained from test beam measurements of all four DUT segments (labeled as TB) and \texttt{Allpix-Squared} simulations (labeled as Sim). Efficiency points are fitted using Function~\ref{eq:eff_fit} up to approximately 5.5 fC as including the lower efficiency points worsens the fit quality in the critical 50 \% efficiency region.}
	\label{fig:segment_eff}
\end{figure}

Full simulations of the ITk R2 module positioned perpendicularly to the beam were performed and compared to the presented experimental results.
The efficiency curve obtained from the simulations is similar to the curves obtained from measurements of all DUT segments, however a difference in the curve slope is observed, as shown on the left panel of Figure~\ref{fig:segment_eff}.
The leading strip median charge value obtained from the simulations is $(3.591 \pm 0.021)$ fC which is slightly lower than the measured values but consistent with them within the stated uncertainties.
The uncertainty contributions are $(\pm0.0008^\text{fit}\pm0.021^\text{syst.}\pm0.0013^\text{stat.})$ fC.
The fit contribution is smaller than in the measurements as the number of simulated thresholds is higher, leading to a more accurate fit.
The systematic and statistical contributions are likewise significantly smaller due to the simulated strips being calibrated uniformly and with capacitive effects of other components not modelled, which leads to a much narrower distribution of median charges on individual strips.

While detection efficiency represents the primary output, it is ultimately influenced by many effects which are not easily distinguishable and separable.
To obtain better insight into the simulation and measurement discrepancy, it can be beneficial to assess a different performance metric.
Mean associated cluster sizes (the mean number of adjacent strips forming a hit cluster associated to a track) represent a more direct way to analyze charge sharing effects, which could for example hint at incorrectly implemented strip geometry or improper noise or cross talk values.

Cluster sizes obtained from simulations are generally in agreement with experimental data for higher thresholds (above 2 fC), as can be seen on the right panel of Figure~\ref{fig:segment_eff}.
For lower thresholds, where cluster sizes vary across the DUT segments due to different noise characteristics, the simulations underestimate the cluster sizes.
There are two factors contributing to this discrepancy.

First, hits caused purely by noise are not generated in the simulations.
Low thresholds, where the noise hits would dominate, therefore exhibit lower mean cluster sizes compared to measurements.
This could be remedied by developing an alternate digitization module in the simulation framework which can account for noise hits.
Second, at the lowest thresholds (below approximately 0.5 fC), the DUT readout electronics is unable to keep up with the extremely high rate of noise hits.
This effect is not modelled in the simulations.
While very low thresholds are explored during measurements, it is not intended to operate the modules in such a configuration.

\subsection{Angled incidence scans}
\begin{figure}[htbp]
	\centering
	\includegraphics[width=0.48\textwidth]{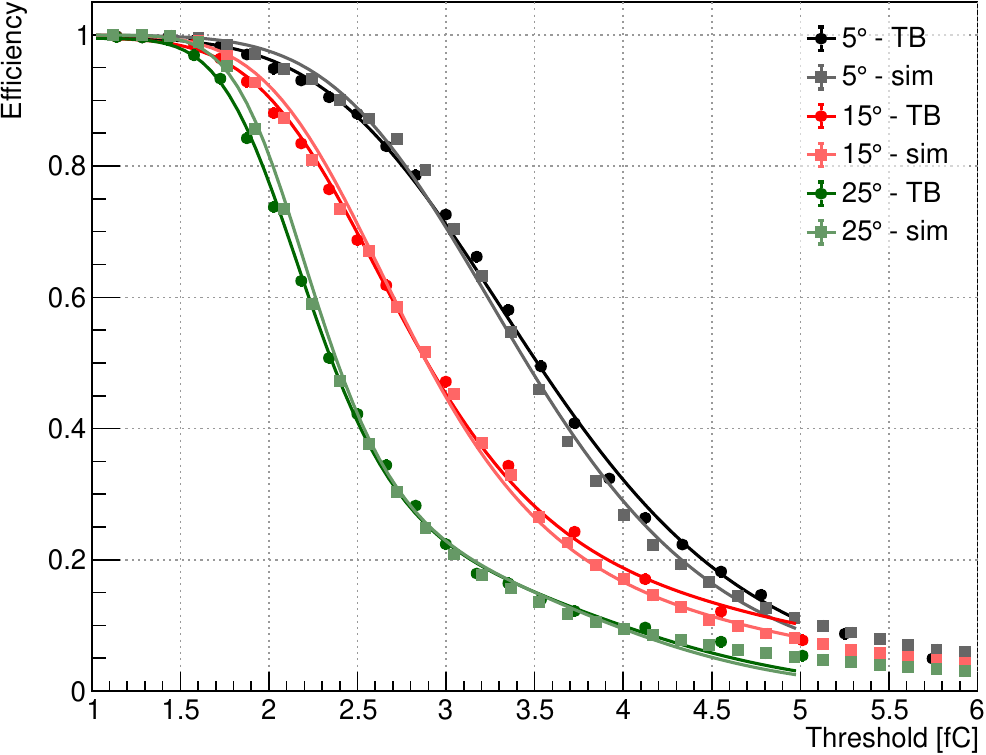}\hskip0.5cm
 	\includegraphics[width=0.48\textwidth]{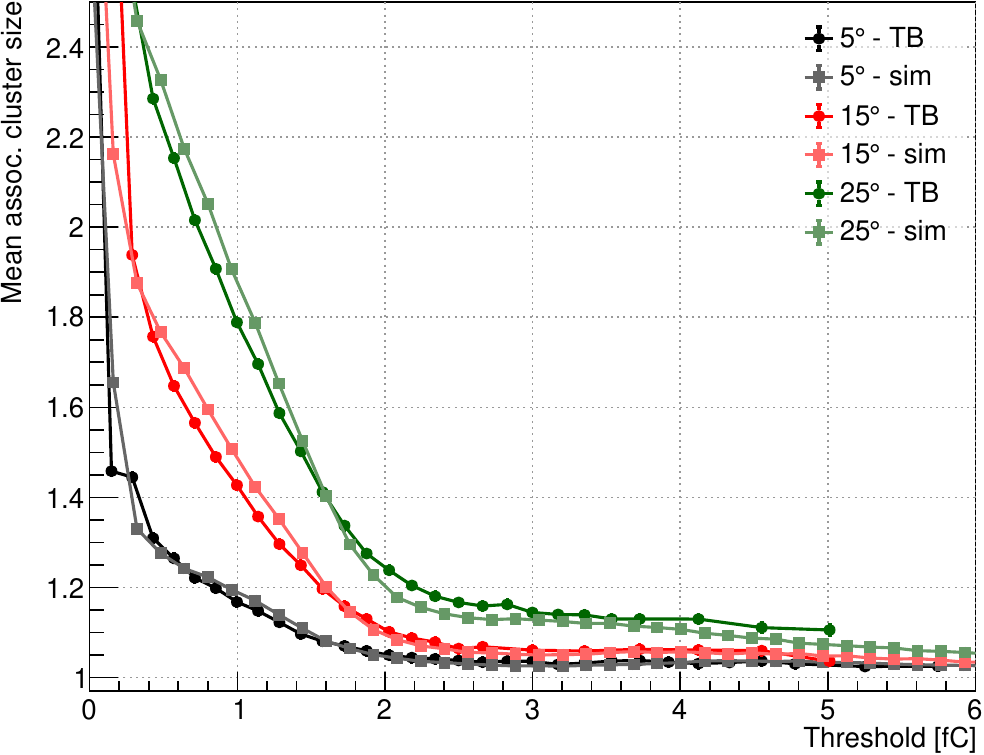}
	\caption{Detection efficiency (left) and mean associated cluster size (right) as a function of the charge threshold for 5$^\circ$, 15$^\circ$ and 25$^\circ$ beam incidence angles, obtained from test beam measurements (labeled as TB) and \texttt{Allpix-Squared} simulations (labeled as sim). Efficiency points are fitted using Function~\ref{eq:eff_fit} up to 5 fC as including the lower efficiency points worsens the fit quality in the critical 50 \% efficiency region.}
	\label{fig:angles_eff}
\end{figure}

Analysis of the angled incidence scans, performed on the s33 segment, shows that the detection efficiency starts decreasing at lower charge thresholds and more rapidly for larger beam incidence angles due to increased charge sharing among the strips, as shown on the left panel of Figure~\ref{fig:angles_eff}. 
The angled beam incidence resulted in the decrease of the leading strip median charge from $(3.82 \pm 0.24)$\,fC for perpendicular beam incidence to $(2.36 \pm 0.29)$\,fC for the 25$^\circ$ beam incidence angle (a~decrease by 38.2 \%) and of the signal-to-noise ratio from $35.9\pm2.3$ to $22.2\pm2.7$.
The stated uncertainties were obtained using the same procedure as in the segment scans analysis.
The operating windows became narrower for larger beam incidence angles as a result of the impacted efficiency, decreasing from 1.47 fC for perpendicular beam incidence to 1.08 fC for the 25$^\circ$ beam incidence angle.
The leading strip median charge values obtained from simulations and measurements, as well as signal-to-noise values and operating window widths, are listed for all beam incidence angles in Table~\ref{tab:angles}.

\begin{table}[hbtp]
    \centering
    \caption{Performance parameters of the s33 segment of the DUT for various beam incidence angles. Listed are leading strip median charges obtained from measurements (labeled TB) and simulations (labeled sim.).}
    \begin{tabular}{c|c|c|c|c}
         \makecell{Incidence\\ angle [$^\circ$]} & \makecell{Median charge\\ (sim.) [fC]} & \makecell{Median charge\\ (TB) [fC]} & S/N & \makecell{Operating window\\ width [fC]} \\ \hline
         0 & $3.591 \pm 0.021$ & $3.82 \pm 0.24$ & $35.9 \pm 2.3$ & 1.47 \\
         5 & $3.440 \pm 0.031$ & $3.52 \pm 0.18$ & $33.1 \pm 1.7$ & 1.23 \\
         10 & $3.169 \pm 0.023$  & $3.29 \pm 0.18$ & $30.9 \pm 1.7$ & 1.15 \\
         15 & $2.893 \pm 0.035$  & $2.92 \pm 0.24$ & $27.4 \pm 2.2$ & 1.12 \\
         20 & $2.653 \pm 0.025$  & $2.62 \pm 0.23$ & $24.7 \pm 2.2$ & 1.11 \\
         25 & $2.411 \pm 0.034$  & $2.36 \pm 0.29$ & $22.2 \pm 2.7$ & 1.08 \\ \hline
    \end{tabular}
    \label{tab:angles}
\end{table}

Efficiency curves obtained from the test beam measurements and \texttt{Allpix-Squared} simulations match reasonably well for the respective beam incidence angles, as shown on the left panel of Figure~\ref{fig:angles_eff}.
Similarly to the perpendicular incidence results, a difference in the slope of the efficiency curves is observed, particularly for the 5$^\circ$ beam incidence angle.
The agreement seems to improve with the increasing incidence angle.
Leading strip median charge values obtained from simulations and measurements for each studied incidence angle are in agreement within their respective uncertainties.

The increased charge sharing for larger incidence angles leads to larger mean associated cluster sizes.
This effect is additionally much stronger at lower charge thresholds (up to approximately 1.5 fC), as is shown on the right panel of Figure~\ref{fig:angles_eff}. 
At these thresholds, the simulations tend to overestimate the cluster sizes, with this effect being more prominent as the incidence angle increases.
Such a result was not observed for perpendicular incidence, where simulations underestimated the cluster sizes compared to measurements, and the cause of this opposite effect is currently not well understood.
At higher thresholds (above 2 fC) the agreement between the simulations and measurements has improved.

\section{Summary}
The analysis of threshold scans taken in June 2022 at the DESY II test beam facility has shown that the measured ATLAS ITk R2 strip module exhibits a wide range of charge thresholds where performance requirements of the ITk collaboration on detection efficiency and noise occupancy are comfortably met.
The individual segments of the module behave fairly consistently in terms of key performance parameters -- detection efficiency, noise occupancy and signal-to-noise -- although some differences were observed among the segments, presumably caused primarily by the presence of the hybrid and the power board on two of the segments.

Angled threshold scans, where the beam was impacting the module at a non-perpendicular angle, showed that the detection efficiency and the amount of signal on the leading strip start decreasing at lower charge thresholds due to increased charge sharing among the strips.
Operating windows were still found for each beam incidence angle, they however become narrower due to the impacted efficiency as the angle increases.
The average cluster sizes become larger with the increasing beam incidence angle; this effect is more prominent at low charge thresholds.

Accompanying computer simulations of the same experimental setup were performed in the \texttt{Allpix-Squared} framework. 
The framework has recently been extended to enable simulations of ITk strip end-cap modules, accurately recreating their complex radial geometry.
The results show the simulations can be used to predict the median charge on the leading strip, both for perpendicular and angled beam incidence.
However, differences between the experimental data and simulations are observed in the shape of the efficiency curves and in mean cluster size values at low charge thresholds.
These deficiencies represent a valuable feedback for further tuning and refinement of the simulation and the implementation of strip detectors therein.

\bibliographystyle{JHEP}
\bibliography{biblio.bib}

\acknowledgments
The measurements leading to these results have been performed at the Test Beam Facility at DESY Hamburg (Germany), a member of the Helmholtz Association (HGF).
The authors gratefully acknowledge the support from the project IGA\_PrF\_2024\_004 of Palacky University.
This work was supported by the European Structural and Investment Funds and the Ministry of Education, Youth and Sports of the Czech Republic via projects LM2023040 CERN-CZ and FORTE - CZ.02.01.01/00/22\_008/0004632.

\section*{Data availability}
The data used to generate plots and tables presented in this document have been made accessible via the institutional data repository of the Czech Academy of Sciences ASEP and is publicly available at \url{https://doi.org/10.57680/asep.0602834}.
\end{document}